
\input harvmac

\Title{HUTP-92/A070}
{\vbox{\centerline{Free field resolution for nonunitary representations of}
	\vskip2pt\centerline{N=2 SuperVirasoro }}}

\centerline{Vladimir Sadov} \bigskip\centerline{Lyman Laboratory of
Physics} \centerline{Harvard University}\centerline{Cambridge, MA 02138}
\centerline{and}\centerline{L.~D.~Landau Institute for Theoretical Physics,
Moscow} 
\vskip .3in

We study N=2 SuperVirasoro SCFT for the generic value of the central charge.
The main tool is the nonstandard bosonisation suggested in
\ref\rSeBGR{B.~Gato-Rivera, A.~Semikhatov {\it Phys.~Letts.}{\bf B293} (1992)
72},\ref\rBLNW{M.~Bershadsky, W.~Lerche, D.~Nemeshansky, N.~Warner
{\it N=2 Extended superconformal structure of Gravity and W Gravity coupled to
Matter}  HUTP-A034/92},
\ref\rRoz{L.~Rozansky  {\it a letter to M.~Bershadsky}, 1989}. The free field
resolutions for the irreducible representations are obtained; the characters of
these representations are computed. The quantum hamiltonian reduction from the
Kac-Moody $\widehat{sl}_k(2|1)$ to N=2  $SVir$ is constructed.

\Date{12/92}

\vfill\eject

\newsec{Introduction}

In dealing with N=2 SCFT, we usually restrict ourselves to the study of of the
unitary theories. One of the reasons is that the representation theory of N=2
SuperVirasoro ( from now on called $SVir$) is much understood for the unitary
case only  \ref\rUniMSS{ G.~Mussardo, G.~Sotkov, M.~Stanichkov {\it
Int.~J.~Mod.~Phys}{\bf A4} (1989) 1135},\ref\rUniIto{ K.~Ito, {\it Nuclear
Physics}{\bf B332} (1990) 566}   . The remarkable feature of these models is
the existence of chiral rings  \ref\rRin{ W.~Lerche, C.~Vafa, N.~Warner {\it
Nuclear Physics}{\bf B324} (1989) 427}  .

Recently it was found  \rSeBGR ,\rBLNW , that N=2 superconformal algebra plays
an important role in 2d gravity coupled to matter. In fact, any noncritical (as
well as critical) string model possesses this symmetry. It is important, that
the {\it unitary} string string theory considered as an N=2 SCFT is
(generically) {\it nonunitary}.\foot{
Though in some interesting cases {\it it is} unitary  \ref\rMuVa{S.~Mukhi,
C.~Vafa {\it Two dimensional Black hole, c=1 Non-Critical Strings and a
Topological Coset Model. } preprint  HUTP-A?92},\ref\rBeOVa{M.~Bershadsky,
S.~Cicotti, H.~Ooguri, C.~Vafa {\it  } preprint  HUTP-93/A??}, already in the
simplest example of (Vir-)minimal matter coupled to gravity it is not.}
Therefore it is interesting to understand such theories better.

In the second section of this paper we introduce the free field resolutions of
all highest weight representations of N=2 $SVir$ using a "new" bosonisation of
\rSeBGR ,\rBLNW ,\rRoz . For the most degenerate representations (
parameterized by three integer numbers ), the corresponding ( Felder type )
BRST complex prove to be one-sided infinite. There are also three noncountable
families of the less degenerate  representations (having only one singular
vector in the Verma module).

In  \rBLNW , it was suggested, that there exists a hamiltonian reduction from
the Kac-Moody $\widehat{sl}_k(2|1)$ to N=2 $SVir$ . It was shown that this is
the case in the classical limit ( $k\rightarrow \infty$ ). Appendix deals with
the quantum version of the reduction. The resolution of the vacuum module
obtained in the Sec.2 gives us a characterization of the embedding of N=2
$SVir$  into the chiral algebra of the free fields, defined by the bosonisation
formulas. Using this characterization we show that the quantum reduction exists
indeed.

Finally, it should be noted that we work with N=2 $SVir$  in a basis, naturally
appearing in a context of the 2d quantum gravity. In this basis, the algebra is
called "twisted", although it is the same $SVir$ .

\newsec{The free field resolutions}

In this section we consider the bosonisation rules, associated with the used in
   \rSeBGR ,\rBLNW  presentation of N=2 $SVir$ . Exactly the same bosonisation
was introduced earlier by L.~Rozansky  \rRoz . Still we prefer to start with
\rBLNW , because it makes the whole construction more transparent. So suppose
we have a system consisting of the Virasoro algebra ($Vir$)
--- a matter sector, a free bosonic field $\phi$ with the background charge
($Heis$) --- a Liouville sector and a pair of fermions $b,c$ of spins 2,-1
($Clif$) --- the diffeomorphism ghosts ( in the brackets are the names of the
corresponding chiral algebras). We require the total central charge be equal to
zero. Then the currents
\eqn\eNtwo{\eqalign{&J(z)=:cb:+2\alpha _-\partial \phi \cr
&G^+(z)=:c[T_{Vir}+T_\phi +{1 \over 2}T_{bc}]:-2\alpha _-\partial (c\partial
\phi ) +{1 \over 2}(1-2\alpha _-^2)\partial ^2c \cr
&G^-(z)=b(z) \cr
&T=T_{Vir}+T_\phi +T_{bc} \cr
&T_\phi =-{1 \over 4}:(\partial \phi )^2:+\beta _0 \partial ^2 \phi \cr
}}

satisfy the OPE of (twisted) N=2 $SVir$  chiral algebra. In fact these formulas
give the embedding of N=2 $SVir$  into the tensor product of three other chiral
algebras
$Vir \otimes Heis \otimes Clif$. It should be stressed that there is such
embedding already for the Virasoro matter, with no bosonisation at all. It
describes a hidden N=2 superconformal symmetry of any string model. But our aim
here is to use the map above to understand better the representation theory of
N=2 $SVir$  itself, rather than the structure of the string models.

To do so, it is convenient to bosonize the matter by the free field $X$ with
the background charge $\alpha _0$. In other words we embed $Vir$ into the
Heisenberg algebra generated by $\partial X$ which we denote by $Heis'$ to
distinguish it from the Liouville $Heis$. Substituting the bosonized matter
stress energy
\eqn\ebos{\eqalign{&T_{Vir}(z) ={1 \over 4}:(\partial X )^2:+\alpha _0 X \cr
&\beta _0^2-\alpha _0^2=1, \ \ \alpha _{\pm}=\alpha _0 \pm \beta _0 \cr
&c_{Vir}=1-24\alpha _0^2 \cr
}}

into \eNtwo
we finally obtain the bosonisation prescriptions we are going to use.
Unlike the standard bosonisation  \rUniIto, \rUniMSS   , the formulas for
$G^+(x)$ and $G^-(z)$ are very asymmetric. It is a whole story at the level of
chiral algebras.

For the representations, we take a Fock space $F_{\alpha \beta}=F_{\alpha}
\otimes F_{\beta} \otimes F_{gh}$. $F_{\alpha}$ and $F_{\beta}$ here are the
standard Fock modules of $Heis'$ and $Heis$  with vacuums $|\alpha >$ and
$|\beta >$  respectively and $F_{gh}$ is a ghosts Fock space (a $Clif$ Verma
module) with the vacuum vector $|0>$ annihilated by
\eqn\evac{c_n|0>=0 \ n > 1, \ \ b_n|0>=0 \ n>-2 }

In $F_{gh}$ we take a vector $|0>_{phys}=c_1|0>$ and define the N=2 vacuum as
$\Omega =|\alpha > \otimes |\beta > \otimes |0>_{phys}$. This procedure is well
known in string theory. Here we use it to endow the free field Fock space
$F_{\alpha \beta}$ with a structure of a highest weight N=2 $SVir$  module:
\eqn\eNtwovac{\eqalign{&L_n\Omega =J_n\Omega =G_n^-=0, \ n\geq 0 \cr
&G_n^+\Omega =0, \ n>0 \cr
&L_0\Omega =(-1+\alpha (\alpha -2\alpha _0)-\beta (\beta -2\beta _0)) \cr
&J_0\Omega =(1+\alpha _-\beta ) \cr
}}

There are two {\it screening operators} in our bosonisation. One of them is
just $E=\oint :e^{\alpha _+X(z)}:$. It comes from the bosonisation of the $Vir$
matter. The other one is $F_1=\oint :e^{-{\alpha _+ \over 2}(X(z)+\phi (z))}:$
\rBLNW,\rRoz,\ref\rDots{Vl.~Dotsenko {/it Mod.~Phys.~Lett.}{\bf A7} (1992)
2505}. It is fermionic and local to itself:\foot{
It is interesting that this operator first has appeared in  \rRoz  . Then it
was rediscovered in  \rDots   where it was used to furnish maps between ground
states of c=1 string}
\eqn\efnilp{F_1^2=0 }

When taken together, $E$ and $F_1$ form a quantum superalgebra
$u_q(n_+(sl(2|1)))$ with $q=e^{\pi i \alpha _+^2}$. Namely they satisfy the
Serre relation:
\eqn\eSerr{E^2F_1-(q+q^{-1})EF_1E+F_1E^2=0}

(As usual, the left hand side of \eSerr is to be understood as a part of some
formal polynomial in screenings acting on the appropriate state, see below.)

We will use $E,F_1$ to construct a family of maps between the Fock spaces
$F_{\alpha \beta}$.  Any  such maps from the Fock space $F_{\alpha \beta}$  can
be presented by the polynomial in generators $e,f$ of $u_q(n_+(sl(2|1)))$
which is, by the definition, just the element of  $u_q(n_+(sl(2|1)))$. The
correspondence between the polynomials and the maps is given by the following
integral representation. On the monomials $e_1 ... fe_{i+1}...fe_{j+1}...e_n$
it is defined as the integral
\eqn\eintrep{\eqalign{&\oint_{z_1}\cdots \oint_{x_n}:e^{\alpha _+X(z_1)}:\ldots
:b(z_i)e^{-{\alpha _+ \over 2}(X(z_i)+\phi (z_i))}:\ldots :b(z_j)e^{-{\alpha _+
\over 2}(X(z_j)+\phi (z_j))}: \cr
&\ldots :e^{\alpha _+X(z_n)}::c(0)e^{\alpha X(0)+\beta \phi (0))}: \cr
}}
over the nested set of contours on the complex plane $|z_1|>\cdots |z_n|$ going
from $x_0$ to $x_0$ around 0, where $x_0$ is an auxiliary point on the complex
plane. Then the definition is continued to all polynomials by linearity. To be
well defined, the integral representation above should not depend on the
position of the auxiliary point $x_0$  \ref\rFaZa{V.~Fateev, A.~Zamolodchikov
{\it Nuclear Physics}{\bf B280} (1987) 644}   . This condition severely
restricts the choice of the polynomials, for the fixed $F_{\alpha \beta}$
excluding all but a finite number of them. Morover, the condition that there
exists a {\it nontrivial} map, restricts also the choice for $({\alpha
,\beta})$

Using the Serre relation \eSerr it is easy to see, that monomial in $e,\ f_1$
can have no more than two $f_1$ factors. Look first at the case when the
polynomial has the order exactly two in $f_1$. For the fixed order in $e$,
there is only one such --- $e^nf_1ef_1$. Let us think of \eintrep as of its
representation. Then {\it if} the corresponding map existed, the integral would
be well defined and the integrated function be single valued. But now we note
that this function is in fact antysymmetric in $z_i$ and $z_j$ ( because of two
$b$'s insertions) so the integral equals zero, i.e. the map we started with was
in fact trivial. Hence we should deal only with the linear in $f_1$
polynomials.

First let us consider three examples of nontrivial maps, which can readily be
checked by looking at \eintrep .
To the monomial $e^n$ there corresponds a nontrivial map if and only if $\alpha
= \alpha _+{1-n \over 2}+\alpha _-{1-m \over 2}$, it is a well known fact from
the representation theory of the Virasoro algebra. There is a nontrivial map
corresponding to  $f$ if and only $\alpha -\beta=-\alpha _-k$. Then, if
$\alpha +\beta=-\alpha _-(k+1)+\alpha _+$ the combination $x_{\alpha}ef_1-f_1e$
with
\eqn\ecfc{x_{\alpha}={q^{2\alpha _-\alpha}-q^{-2\alpha _-\alpha}   \over
q^{2\alpha _-\alpha +1}-q^{-2\alpha _-\alpha -1}  }={[2\alpha _-\alpha ]_q
\over [2\alpha _-\alpha +1]_q}} is the map.

We call a map {\it primitive}, if it cannot be obtained as a composition of two
nontrivial maps. It is clear, that if we know all primitive maps, we know
everything. The results of the manipulations with the integral
representation\foot{ This can also be observed from the formulas for the
singular vectors in the Verma modules of $u_q(sl(2|1))$.} \eintrep can be put
in the following nice form.

{\it There are only three types of the primitive maps, listed above.}

Let us consider the oriented graphs (the diagrams) whose vertices are the Fock
spaces $F_{\alpha \beta}$ and arrows are the primitive maps between them,
belonging to our family. It is important that the resulting map between two
vertices does not depend on the path we choose to connect them. From now on
(unless explicitely stated otherwise) let us restrict ourselves to case when
$\alpha _+^2$  is not a rational number. The more complicated case of rational
$\alpha _+^2$  can be treated by exactly the same technique and will be
considered elsewhere  \ref\rFuru{V.~Sadov {\it Fusion rules in N=2 nonunitary
SCFT} preprint HUTP-93/A??, in preparation}.

Then every $F_{\alpha \beta }$ belongs to the diagram of one of the four types,
shown in Fig.1.\foot{ We mark the Fock space $F_{\alpha \beta }$ by the type of
the $u_q(sl(2|1))$ Verma module it corresponds to and the irreducible
representation $L_{\alpha \beta }$ by the type of the Fock space having the
same highest weight.}

If we knew that the maps which can be expressed in terms of screenings were
{\it all} the maps between the Fock spaces, we could conclude that we know
completely the primitive (i.e. singular and cosingular) vectors of the latter.
Then a more or less straightforward construction below should provide us with
the resolutions we are after. In fact, we cannot {\it prove} that we have
described all the maps, so we have to {\it assume} that. The general problem of
formulating exactly the correspondence between the quantum groups
($u_q(sl(2|1))$ in our case) and the chiral algebras is still open (but cf.
\ref\rLuKa{D.~Kazhdan, G.~Lusztig {\it Tensor structures arizing from affine
Lie algebras I,II} preprints}  !), so one should think of this assumption as of
(very well grounded) hypothesis.

We are ready now to construct the free field resolutions of the irreducible
representations of N=2 $SVir$\foot{ In fact, the construction goes parallel to
that of the BGG resolution for $u_q(sl(2|1))$.}. By such resolution we mean the
complex of the free field Fock spaces with the cohomology being nontrivial only
in the zero degree where it is represented by the irreducible representation.
Having a complex we compute its Euler characteristics (character valued, as
usual) which turns to be a character of the irreducible $L_{\alpha \beta }$. It
is convenient to deal with the normalized characters
\eqn\eintrep{\eqalign{
&\tilde{\chi}_{\alpha \beta}={\chi _{\alpha \beta} \over \chi (F_{\alpha ,
\beta})}\ , \ \chi (F_{\alpha , \beta})=Tr_{F_{\alpha ,
\beta}}(q^{L_0}x^{2J_0}) \cr
}}

{\bf Case I}. $(\alpha , \beta )$ is generic, the module $F_{\alpha \beta }$ is
irreducible. The corresponding complex is therefore trivial,
$\tilde{\chi}_{\alpha \beta}=1$.

{\bf Case II.}
\eqn\econdII{\alpha _{n\ m}= \alpha _+{1-n \over 2}+\alpha _-{1-m \over 2} }
$\beta$ is generic. This case essentially reduces to the well known theory for
the Virasoro algebra. The map $E^n$ is surjective, its kernel is a submodule in
${F_{\alpha  \beta}}$ generated by the highest weight vector. It gives $L_{m\
n}(Vir)\otimes F_{\beta}\otimes F_{gh}$. By our assumption this is the only map
{\it from} $F_{\alpha  \beta}$ and the only map {\it to} $F_{\alpha +n\alpha
_+\ \beta}$; there is no maps {\it to} $F_{\alpha  \beta}$ or {\it from}
$F_{\alpha _+\alpha _+\ \beta}$. Therefore the submodule in $F_{\alpha  \beta}$
generated by the highest weight vector is irreducible. The quotient of
$F_{\alpha  \beta}$ by this submodule is also irreducible and coincides with
$F_{\alpha +n\alpha _+\ \beta}$. The character is
\eqn\echarII{\tilde{\chi}_{\alpha \beta}=1-q^{nm}}
This was the typical example of the argument using our basic assumption. It is
always implied below.

{\bf Case $III_-$.}
\eqn\econdIIIminus{\alpha -\beta =-\alpha _-k }
When $k \geq -1$ the map $F_1$ sends the highest weight vector of ${F_{\alpha ,
\beta}}$ to a nonzero element, which generate in $F_{\alpha -{\alpha _+\over
2}, \beta -{\alpha _+\over 2}}$ a proper submodule $SF_{\alpha -{\alpha _+\over
2}, \beta -{\alpha _+\over 2}}$. There is no other maps into $F_{\alpha
-{\alpha _+\over 2}, \beta -{\alpha _+\over 2}}$, so $SF_{\alpha -{\alpha
_+\over 2}, \beta -{\alpha _+\over 2}}$  is the only proper submodule.
Therefore it must coincide with the kernel of the map $F_1$ from $F_{\alpha
-{\alpha _+\over 2}, \beta -{\alpha _+\over 2}}$ to $F_{\alpha -{\alpha _+},
\beta -{\alpha _+}}$. This means that the diagram $III_-$ is {\it exact} ---
the image of the ingoing arrow coincide with the kernel of the outgoing arrow.
In particular it imply that the composition of two arrows is zero. But this is
certainly true as it is just $F_1^2=0$ \efnilp . Althogh trivial, this fact
should be considered as a consistency check for our assumption.

The diagram $III_-$ has already a natural structure of the complex. The graded
components are just the Fock spaces at the vertices and the differentials are
given by the arrows. This complex is infinite in both directions. We now it is
exact, so its cohomology is trivial. Therefore another (less trivial)
consistency check is to compute its Euler character to make sure it equals
zero.  Easy to see, it is zero indeed. To obtain the resolution of $L_{\alpha
\beta }$ one cuts the diagram by the arrow going {\it from} $F_{\alpha \beta }$
to obtain two complexes with the equal cohomology (so there are two resolutions
in fact), one can use either of them. The character is
\eqn\echarIIIminus{\eqalign{
&\tilde{\chi}_{\alpha \beta}={1 \over 1+x^{-1}q^{k+1}} \cr
}}
We see that for $k \geq -1$ the formula \echarIIIminus can naturally be
interpreted as a character of the representation with the highest weight
$(\Delta _{\alpha \beta},q_{\alpha \beta})$. But for $k <-1$ the identical
transformation
\echarIIIminus $\rightarrow {xq{-k-1} \over 1+xq^{-k-1}}$ shows that the
character we compute now correspond to the weight  $(\Delta _{\alpha +{\alpha
_+ \over 2}, \beta +{\alpha _+ \over 2}},q_{\alpha +{\alpha _+ \over 2},\beta
+{\alpha _+ \over 2}})$. The reason for this phenomena is simple. For $k <-1$
the map $F_1$ kills the highest weght vector of $F_{\alpha \beta }$ and sends
some vector $w \in F_{\alpha \beta }$ to the highest vector of $F_{\alpha
-{\alpha _+\over 2}, \beta -{\alpha _+\over 2}}$. Hence each Fock space has one
cosingular vector $w_{\alpha \beta}$ and the irreducible representation is a
{\it submodule} of $F_{\alpha \beta }$ generated by the highest weight vector
of the Fock space\foot{Compare to $k \geq -1$ when the irr. rep. was a {\it
quotient } of the Fock space.}. Now, to obtain a resolution of $L_{\alpha \beta
}$ we should cut the diagram $III_-$ by the arrow coming {\it into} $F_{\alpha
\beta }$. The character is
\eqn\echarIIIplus{\eqalign{
&\tilde{\chi}_{\alpha \beta}={1 \over 1+xq^{-k-1}} \cr
}}

{\bf Case $III_+$}
\eqn\econdIIIpluplu{\alpha +\beta=\alpha _-(k+1)+\alpha _+}

One can repeat everything that have been said about $III_-$. The only subtlety
here is to check that the composition of two consequitive maps in the diagram
is zero. The reader should convince oneself it is true using \eSerr ,\ecfc and
simple $q$-polynomial identities. The formulas for the characters are given by
the same formulas \echarIIIminus ,\echarIIIplus .

{\bf Case $IV_-$} --- the conditions II and $III_-$ ($IV_-$) are met
simultaneously.
\eqn\econdIV{\eqalign{
&\alpha _{nm}= \alpha _+{1-n \over 2}+\alpha _-{1-m \over 2} \cr
&\beta _{nmk}= \alpha _+{1-n \over 2}+\alpha _-{1-m+2k \over 2} \cr
}}
 We denote $F_{\alpha _{nm}\beta _{nmk}}$ by $F_{nmk}$. We know everything
already about the maps in the diagram. First consider the resolution of the
representation with $n\geq 0$, i.e. belonging to the left column in Fig.1. To
obtain a resolution we should again cut the diagram by the horizontal line
crossing the arrow  {\it above} $(\alpha \beta )$ for $k\geq -1$  or {\it
below} $(\alpha \beta )$  for $k<-1$. Holding the upper half end up with a
"ladder" shown in the Fig.2.
The structure of the complex $\{C^r, d_{(r)}\}_{r\geq 0}$is given by\foot{
It is a good exercise to check, using the Serre relation, that
$d_(r+1)d_(r)=0$.}
\eqn\ecompIV{\eqalign{
&C^0=F_{n+1\ m\ k},\ C^r=F_{n+1+r\ m\ k}\oplus F_{-(n+r)\ m\ k} \cr
&d_{(0)}=E^{n+1}\oplus F_1, \ d_{(r)}= \pmatrix{&F_1 &0 \cr
					      &E^{n+1+r} &x_{n+r}EF_1-F_1E \cr}
}}
\eqn\ecreq{x_{l+1}=(q+q^{-1})-{1\over x_l},\ \ x_0=q+q^{-1}}
 The character is
\eqn\echarIV{\eqalign{
&k\geq -1 \ \ \tilde{\chi}_{\alpha \beta}={1-q^{m(n+1)}+q^{m-k-1}(1-q^{mn})
\over (1+x^{-1}q^{k+1})(1+xq^{m-k-1})} \cr
&k<-1 \ \ \tilde{\chi}_{\alpha \beta}={1-q^{mn}+q^{m-k-1}(1-q^{m(n-1)}) \over
(1+xq^{-k-1})(1+xq^{m-k-1})} \cr
}}
Now about the representations with $n<0$ (in the right column in Fig.1). The
image of the map $E^n$ in such Fock space is generated by the highest weight
vector. Thus we infer that all these modules are of type $III_+$ with
$k'=m-k-1$.

{\bf Case $IV_+$} --- the conditions II and $III_+$ are met simultaneously.
\eqn\econdIV{\eqalign{
&\alpha _{nm}= \alpha _+{1-n \over 2}+\alpha _-{1-m \over 2} \cr
&\beta _{nmk}= \alpha _+{1+n \over 2}+\alpha _-{-1+m+2(k+1) \over 2} \cr
}}
One just repeats what was said about $IV_+$ (probably it is better to take a
bottom half of the cut diagram to construct a resolution of the representations
with $n>0$ in this case). The representations with $n<0$ are of type $III_-$.
The formulas for the characters \echarIV are applicable.

Thus, using the basic assumption we come up with the consistent results for the
resolutions and characters\foot{
An expert reader must have noticed already that $I-IV$ are equivalent to saying
that there is always a resolution of N=2 $Svir$ ir.~rep. by the modules
($Vir$-ir.~rep.)$\otimes (Fock_{\phi})\otimes (Fock_{gh})$. It is in this form
that $I-IV$ are generalized to the rational $c$.}. Still it is desirable to
have additional evidence these results are true. Fortunately, there is another
way to prove that we have a resolution, but it works only for the generic
values of $\alpha _+^2$and only for some representations. Following
\ref\rFeFr{B.~Feigin, E.~Frenkel
{\it Phys.~Letts.~}{\bf B246}(1990) 75}, \ref\rFr{E.~Frenkel
{\it a talk at the Cargese Sumer School}(1991)}  , we consider a classical
limit ($\alpha _+^2 \rightarrow 0$). Rescaling the generators of the Heisenberg
algebras and the screenings we end up with the action of the (classical)
$n_+(sl(2|1))$ just on $(F_{\alpha \beta })_{class}$, if the latter limits
exist. (The problem is that as $\alpha _+^2 \rightarrow 0$, $\alpha _-^2
\rightarrow \infty$, so for example for the $IV$ one should restrict $m$ to 1
to and $k$ to 0 to have the meaningful limits). The BRST complexes above become
just the BGG complexes of $sl(2|1)$, to be more exact, they have the structure
of $Hom(BGG(\alpha ,\beta ), (F_{\alpha \beta })_{class})$. Hence
the cohomology of these complexes is just that of the superalgebra
$n_+(sl(2|1))$ with the coefficients in $(F_{\alpha \beta })_{class})$.
But one can check (it was known for $E_{class}$ and is fairly obvious for
$(F_1)_{class}$) that the action of $E_{class}$,$(F_1)_{class}$ on $(F_{\alpha
\beta })_{class}$ is co-free. It implies that $H^0_{d^{class}}(C)$ --- the
invariants of $n_+(sl(2|1))$ = $(L_{\alpha \beta })_{class}$ and the higher
cohomologies are trivial. Now, using the semicontinuity of the cohomology of
the algebraic family of complexes we conclude that for the generic value of
$\alpha _+^2$ the higher cohomologies of the BRST complexes above are still
trivial and $H^0_{d}(C)$ is still $L_{\alpha \beta }$.

An important representation which has a classical limit and therefore surely
has a free field resolution (of type IV) is a vacuum representation. We have
seen, that it coincides with the intersection of the kernels of two screenings
$E,\ F_1$ acting on the vacuum Fock space $F_{00}$. On the other hand, the
vacuum irreducible representation is a vertex operator algebra (VOA)
\ref\rFLepM{I.~Frenkel, J.~Lepowsky, A.~Meurman {\it Vertex Operator Algebras
and the Monster} Academic Press, 1988}   of N=2 $SVir$ . Therefore, we have
proved that

{\it As a chiral algebra N=2 $SVir$  in the product $Heis' \otimes Heis \otimes
Clif$ is characterized as a centralizer of two screening operators $E,\ F_1$.}

\newsec{Conclusion}

We have considered the quantum group approach to the representation theory of
N=2 $SVir$. From the point of view of the pure mathematics, our main assumption
 is but a hypothesis. However, we have seen that it gives consistent results
and can be vindicated for the generic central charges $c$. It is natural to
conjecture that in fact it holds for all values of $c$. Then for the rational
we can construct the (Felder type) resolutions simply looking at the
representations of $u_q(sl(2|1))$ in the roots of unity. Another obvious
generalization is for the N=2 $W$ algebras.

Then, it is interesting to study the fusion rules of the N=2 SCFT, coresponding
to the representation theory we have described. It apears that it can be done
using essentially the same technique  \rFuru  .

\appendix{Hamiltonian reduction}{}

Now we wish to use the last result of the section 2 to prove that N=2 $SVir$
(as a chiral algebra) can be obtained from the Kac-Moody $\widehat{sl}_k(2|1)$
by the hamiltonian reduction, as it was suggested in  \rBLNW  .
To do this, we consider a BRST reduction complex. We take the chiral algebra
$\widehat{sl}_k(2|1)$
\eqn\essl{\eqalign{
&B^+(z)B^-(0)={1 \over z}H(0)+{k \over z^2},\  H(z)H(0)={2k \over z^2}\cr
&S(z)B_{\pm}(0)=\mp {1 \over z}B_{\pm}(0),\  F_1^+(z)F_1^-(0)={1 \over z}S(0) +
{k \over z^2} \cr
&S(z)S(0)=0,\  F_2^+(z)F_2^-(0)={1 \over z}(S(0)+H(0)) + {k \over z^2} \cr
&H(z)S(0)=-{k \over z^2} \cr
}}

and add to it four ghost-antighost pairs. Two bosonic --- $(\beta ,\gamma )$
and $(\beta ',\gamma ')$
and two fermionic --- $(b,c)$ and $(b',c')$. Then we put the constraints
encoded in the BRST reduction operator
\eqn\eBRST{\eqalign{
&d_{BRST}=\oint \{c'(B^+-1)+\beta (F^+_1-b)+\beta 'F^+_2+c'\beta \gamma '\} \cr
}}
To compute the cohomology (in a fashion similar to that used in  \rFeFr, \rFr
) it is convenient to decompose BRST operator $d_{BRST}=d_0+d_1$
\eqn\espesq{\eqalign{
&d_0=\oint (c'B^+ + \beta F^+_1 + \beta 'F^+_2 + c'\beta \gamma ') \cr
&d_1=-\oint (c'+\beta b) \cr
}}
into two pieces and then use the spectral sequence technique.

The cohomology of $d_0$  can easily be computed. It is generated (as a chiral
algebra) by the currents $c'(z)$, $\beta (z)$, $(b(z),c(z)$ and
\eqn\ecoh{\eqalign{
&\tilde{H}=H+2:b'c':+:\beta '\gamma ':-:\beta \gamma :\cr
&\tilde{S}=S-:b'c':-:\beta '\gamma ': \cr
}}
They are not algebraically independent --- in $H^*_{d_0}$
the identities $[d_0,B^-(z)]=0$, $\{d_0,F_1^-(z)\}=0$
give the relations
\eqn\erel{\eqalign{
&:\tilde{H}c':+(k+1)\partial c'=0 \cr
&:\tilde{S}\beta :+(k+1)\partial \beta =0 \cr
}}
Introducing the new variables ($\alpha _+={1 \over \sqrt{2(k+1)}}$)
\eqn\efreefi{\eqalign{
&X(z)=\alpha _+ \tilde{H}(z) \cr
&\phi (z)=-\alpha _+ (\tilde{H}(z)+2\tilde{S}(z)) \cr
}}
we see that on $E_1^{0 \cdot}$ (by definition this subspace contains no ghosts
$c'(z)$, $\beta (z)$, so it is just $Heis' \otimes Heis \otimes Clif$ of
$X(z)$, $\phi (z)$ and $(b,c)$) the currents $c(z)$, $\beta (z)$ act as the
vertex operators $c'(z)=:e^{\alpha _+X(z_1)}:$, $\beta (z)=:e^{-{\alpha _+
\over 2}(X(z_i)+\phi (z_i))}:$. Thus the cohomology of $d_1$ restricted to
$E_1^{0 \cdot}$  is the the centralizer of these two screenings and therefore
coincides with N=2 $SVir$.
On the other hand, because of the relations
\eqn\etriv{\eqalign{
&c'(z)=[d_1,S(z)] \cr
&\beta (z)=\{d_1,c(z)\} \cr
}}
the cohomology of $d_1$ on the compliment to $E_1^{0 \cdot}$ is zero.

 Finally, {\it the cohomology of the BRST reduction complex we described is
given by N=2 $SVir$}.

\bigbreak\bigskip\bigskip\centerline{{\bf Acknowledgements}}\nobreak

I thank M.~Bershadsky  for bringing to my attention the problem of free field
resolution in the Rozansky-BLNW bosonisation and lots of  illuminating
discussions.

I am grateful to C.~Vafa and E.~Frenkel,
for the interest to this work and useful discussions.

Research supported in part by the Packard Foundation and by NSF grant
PHY-87-14654

\listrefs
\bye